\documentstyle[11pt,newpasp,twoside,epsf]{article}
\def\edcomment#1{\iffalse\marginpar{\raggedright\sl#1\/}\else\relax\fi}
\reversemarginpar

\begin{document}
\title{Unpulsed Optical Emission from the Crab Pulsar}
 \author{Aaron Golden, Andy Shearer}
\affil{The National University of Ireland,
Galway, Newcastle Road, Galway, Ireland}
\author{Gregory Beskin}
\affil{Special Astrophysical Observatory, Russia}

\begin{abstract}

Using the high speed 2-d TRIFFID photometer, we have obtained phase
resolved photometry of the Crab pulsar in $UBV$ that allows us
to flux the unpulsed light curve component. Following de-extinction, 
weighted least-square fitting indicates a power-law exponent of
$\alpha$ = -0.62 $\pm$ 0.49. This is steeper than that reported
for the peak components and its origin remains unclear with respect
to contemporary magnetospheric theory.

\end{abstract}

\section{Introduction}

Optically to date phase-resolved observations of the Crab pulsar have been restricted to 
rudimentary 2-d or single pixel photometers, with
variable temporal accuracy. Successes worthy of note include the detection
of an unpulsed component by Peterson et al. (1978), and strong polarisation
behaviour as a function of phase - most especially for this same unpulsed
component (Smith et al. 1988). True characterisation of this interesting
feature requires multi-band high speed 2-d photometry at $\mu$sec 
resolution, allowing for the acquisition of accurate phase-resolved 
photometry. We outline an analysis of such observations made of the
Crab pulsar using the 
TRIFFID high speed photometer (Golden \& Shearer, 1999).

\section{Technical \& Analytical Overview}

The TRIFFID camera incorporating a MAMA detector 
was used to observe the Crab pulsar over 5 nights in January 1996 on the 6m
telescope in the Russian Caucasus.
Using the Jodrell Crab Ephemeris (Lyne \& Pritchard, 1996), 
photons within specific phase regions were selected
to produce a sequence of phase-resolved images over the pulsar's
light curve. Figure 1 shows such a sequence, and confirms the
reported constant emission. Via standard image reduction techniques, 
the relative fluxes per light curve component per colour band were determined, and
normalized to archival integrated estimates (Percival et al. 1993).
Flux components per colour band are plotted in Figure 1, and
a weighted least squares fit to the unpulsed components indicate
a power-law exponent of -0.6 $\pm$ 0.4, the reference integrated
photometry contributing most of the associated error. 

\section{Discussion \& Conclusions}

We have resolved constant emission from the `off' phase of the Crab
light curve. 
Phenomenologically, the
emission is nonthermal and {\it steeper} than that observed for the
peaks, although similar to that of the bridge component. 
Whether it is directly related to the latter, or is
a consequence of magnetospheric scattering is not clear. 
Further observations of this \& the Vela pulsar may show evidence
for a similar unpulsed component of emission, which as yet
remains inexplicable in terms of contemporary high energy
emission theory. 

\begin{figure}
\plottwo{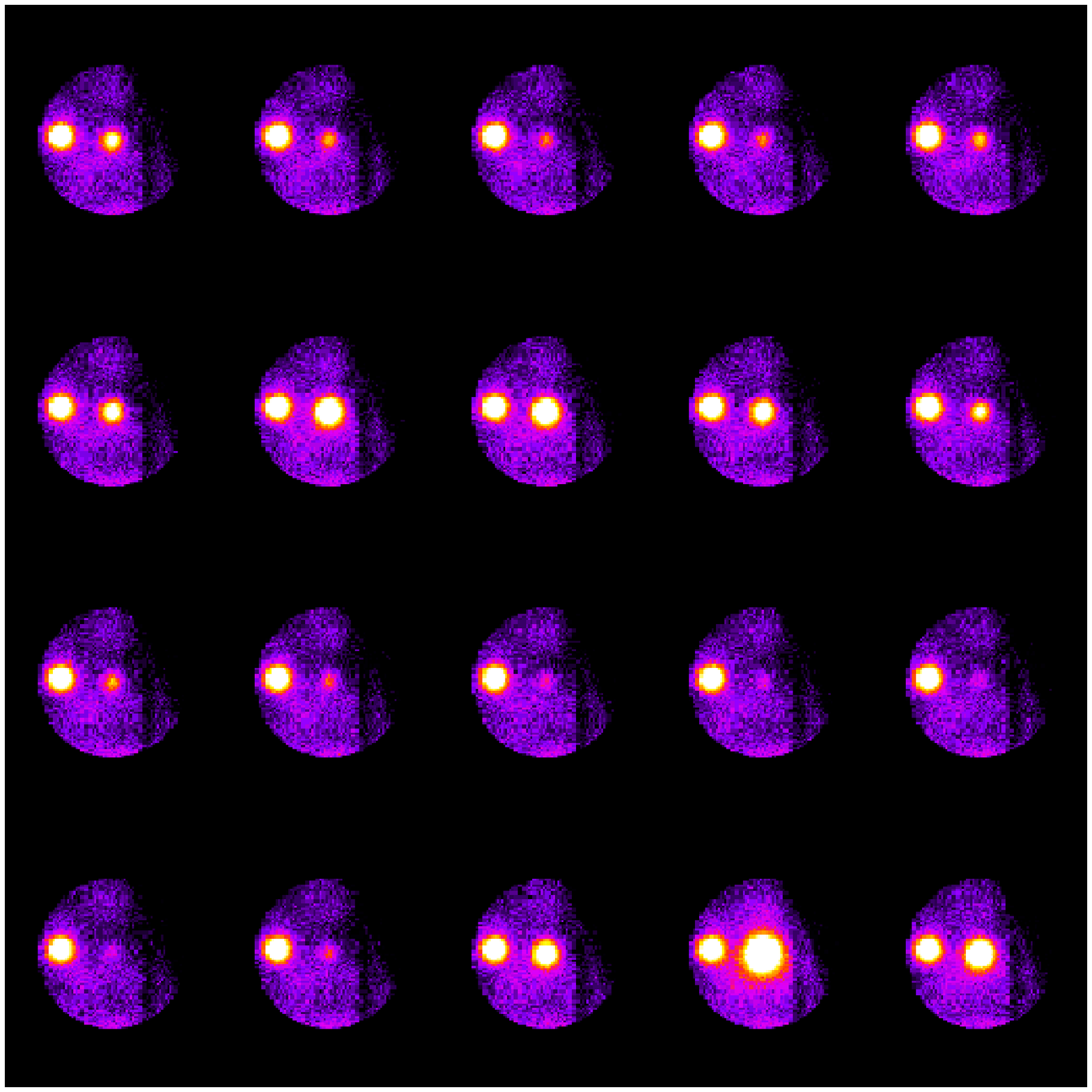}{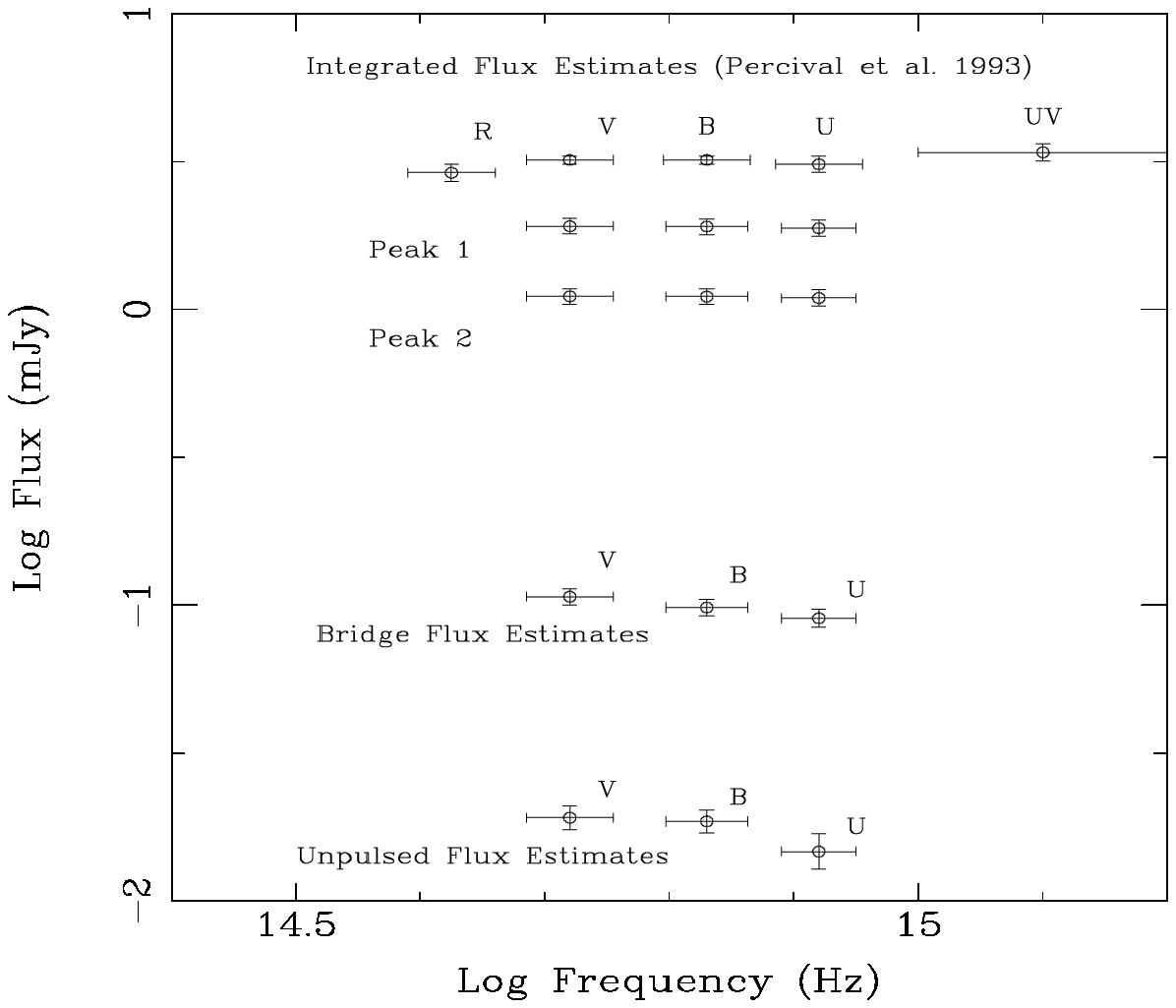}
\caption{\scriptsize {\bf Left:} Mosaic image of one full cycle of the Crab pulsar, via phase-resolved 
photometric
analysis. The light curve moves from top left to bottom right. Note the `on' emission 
throughout. {\bf Right:} Integrated de-extincted flux estimates for the Crab
from Percival et al. 1993 and the derived flux estimates for 
both the peaks, Bridge and unpulsed component of emission 
based on the TRIFFID datasets.}
\end{figure}

\end{document}